\documentclass[prl,twocolumn]{revtex4}
\usepackage{graphicx}

\begin{document}
\title{A Dissipative Tonks-Girardeau Gas of Molecules}

\author{S. D{\"u}rr,$^1$ N. Syassen,$^1$ D.~M. Bauer,$^1$ M. Lettner,$^1$ 
T. Volz,$^{1,}$\footnote{Present address: Institute of Quantum Electronics, ETH-H\"{o}nggerberg, 8093 Z\"{u}rich, Switzerland} \\
D. Dietze,$^{1,}$\footnote{Present address: Institut f\"{u}r Photonik, Technische Universit\"{a}t Wien, Gu{\ss}hausstr. 25-29, 1040 Wien, Austria} 
J.-J. Garc\'{i}a-Ripoll,$^{1,2}$ J.~I. Cirac,$^1$ and G. Rempe$^1$}
\affiliation{$^1$ Max-Planck-Institut f{\"u}r Quantenoptik, Hans-Kopfermann-Stra{\ss}e 1, 85748 Garching, Germany \\
$^2$ Universidad Complutense, Facultad de F\'{i}sicas, Ciudad Universitaria s/n, Madrid 28040, Spain}

\begin{abstract}
Strongly correlated states in many-body systems are traditionally created using elastic interparticle interactions. Here we show that inelastic interactions between particles can also drive a system into the strongly correlated regime. This is shown by an experimental realization of a specific strongly correlated system, namely a one-dimensional molecular Tonks-Girardeau gas.
\end{abstract}

\maketitle

Strong correlations give rise to many fascinating quantum phenomena in many-body systems, such as high-temperature superconductivity \cite{anderson:87}, excitations with fractional statistics \cite{wilczek:82}, topological quantum computation \cite{kitaev:03}, and a variety of exotic behaviors in magnetic systems \cite{auerbach:94}. Such strong correlations are typically the result of a repulsive, elastic interparticle interaction. This makes it energetically unfavorable for particles to be at the same position and thus the wave function tends to vanish at those positions.

In this work we show that inelastic interactions offer an alternative route into the strongly correlated regime. The inelastic collisions also lead to a situation in which the wave function tends to vanish at positions where two particles are at the same position.

This behavior might seem counter-intuitive, but it can be understood in terms of an analogy in classical optics: consider an electromagnetic wave in a medium with refractive index $n_1$ impinging under a right angle onto a surface to another medium with refractive index $n_2$. Fresnel's formula yields that a fraction
\begin{eqnarray}
\label{R}
R = \left|\frac{n_1-n_2}{n_1+n_2}\right|^2
\end{eqnarray}
of the light intensity is reflected. This formula also holds if absorption occurs in the media. In a nonmagnetic medium, dispersion and absorption are expressed by the real and imaginary parts, ${\rm Re}(\chi)$ and ${\rm Im}(\chi)$, of the electrical susceptibility $\chi$, which determines the refractive index $n=\sqrt{1+\chi}$.

In Fresnel's formula, the limit $|n_2|\to\infty$ yields $R\to1$, which is called index mismatch. This result is independent of whether $n_2$ is real or complex. Even for strong pure absorption, ${\rm Im}(\chi_2)\to\infty$, we obtain $R\to1$. In this limit, the light would be absorbed very quickly once inside the medium, but the index mismatch prevents it from getting there.

We add that the result $R\to1$ in the limit of strong loss is quite different from the well-known result $R\approx1$ for light impinging onto a metal. In a metal with negligible Ohmian resistivity, the interaction is purely dispersive, so that $\chi_2$ is purely real. At frequencies below the so-called plasma frequency, a metal has $\chi_2<-1$ so that $n_2$ is purely imaginary, thus causing $R=1$ for real $n_1$ and for any value of ${\rm Im}(n_2)$. This reflection is due to surface charges that build up. They cancel the field inside the metal, which results in a reflection of the wave from the surface. In the static limit, we obtain the well-known result that a static electric field cannot penetrate a conductor because of surface charges. For comparison, for reflection in the limit of strong loss, ${\rm Im}(\chi_2)\to\infty$, we obtain $n_2\approx e^{i\pi/4} \sqrt{|\chi_2|}$ so that ${\rm Re}(n_2)\approx {\rm Im}(n_2)$. Furthermore, $R\approx1$ is only obtained for large ${\rm Im}(\chi_2)$.

These reflection properties of light waves have a one-to-one analogy in the reflection of matter waves. This becomes obvious when defining the refractive index for matter waves obeying the Schr\"odinger equation as \cite{adams:94} $n({\bf x})=\sqrt{1-V({\bf x})/E}$, where $V({\bf x})$ is the potential and $E>0$ is the energy of the particles. One can show fairly easily that the fraction of the particles that are reflected from a potential step at normal incidence is also given by Eq.\ (\ref{R}). Loss of particles is expressed by ${\rm Im}(V)<0$ and strong loss, ${\rm Im}(V_2)\to-\infty$, yields $R\to1$, in full analogy to classical optics.

What we are really interested in is a system, in which the loss is not caused by a static medium, but by inelastic interparticle interactions. The analogy to the case of reflection from the surface to a static medium is clear: once the wave functions of two particles overlapped, they would be lost very quickly. This corresponds to an index mismatch which prevents the wave functions from overlapping. The particles are thus reflected from one another, causing the wave function to vanish at the positions where they would overlap, just like in the case of strong elastic interactions.

An example of a strongly-correlated system that is well known in the field of ultracold gases is a Mott insulator of atoms in an optical lattice \cite{greiner:02}. In this system, strong elastic, repulsive interactions between the particles make it energetically favorable that each site of an optical lattice contains the exact same number of particles. In a recent experiment \cite{volz:06}, we demonstrated that this interaction-induced property of an atomic gas can be maintained when converting atom pairs to molecules. To this end, an atomic Mott isolator was prepared such that it contained exactly two atoms at each site of the central region of the optical lattice. These atom pairs were then associated into molecules \cite{duerr:04} using a Feshbach resonance \cite{marte:02}. In Ref.\ \cite{volz:06} the lattice was so deep that tunneling of molecules was negligible on the time scale of the experiment. The association thus mapped the strong correlations of the interaction-induced atomic Mott insulator to a corresponding quantum state of molecules. But this state was independent of any molecule-molecule interactions.

Studies of the excitation spectrum of this system showed that the molecule-molecule interactions are predominantly inelastic \cite{duerr:06:APS}. In the following, we thus neglect the elastic part of the molecule-molecule interactions. The inelastic character of the collisions arises from the fact that the molecules formed with the Feshbach resonance are in a highly excited ro-vibrational state. If two such molecules collide, it is possible that one of them falls down on the vibrational ladder, thus releasing binding energy into kinetic energy of the relative motion of the colliding particles. The released energy is typically much larger then the trap depth, so that all collision partners are quickly lost from the sample.

\begin{figure}[tb!]
\centering
\includegraphics[width=6cm]{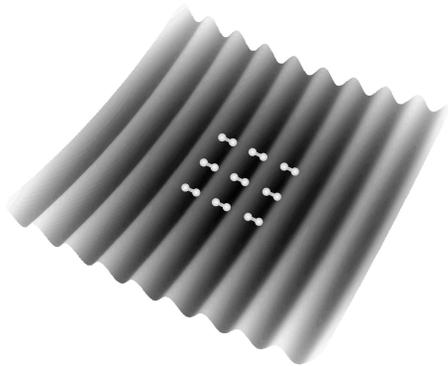}
\caption{\label{fig-scheme}
2D scheme of the 3D experiment in the optical lattice. The lattice depth along one dimension is lowered to zero so that the molecules are free to move along the resulting 1D tubes.}
\end{figure}

A natural question to ask is: What happens to this quantum state if the lattice depth is reduced so much that tunneling of molecules becomes significant? We addressed this question experimentally in Ref.\ \cite{syassen:08} and we discuss the results of this experiment in the following. As hinted above, we find that for sufficiently strong inelastic interactions the system remains strongly correlated. It is well known that a reduction of the dimensionality of the system makes it easier to reach the strongly-correlated regime. We thus choose to lower the depth of the three-dimensional (3D) optical lattice only along one direction, as shown in Fig.\ \ref{fig-scheme}. The strongly correlated gas of repulsively interacting particles in the resulting 1D tubes is known as a Tonks-Girardeau gas \cite{tonks:36,girardeau:60} and has been observed in previous experiments with atoms \cite{paredes:04,kinoshita:04}. Instead of elastic interactions, our experiment relies on inelastic interactions to reach the strongly correlated regime and thus represents a dissipative Tonks-Girardeau gas. Another novel aspect of our experiment is that it realizes the first Tonks-Girardeau gas of molecules.

A characteristic property of the Tonks-Girardeau gas is that the probability to find two particles at the same position is strongly suppressed \cite{gangardt:03,kinoshita:05}. A mathematical expression that captures this property is the pair-correlation function
\begin{eqnarray}
g^{(2)}({\bf x}_1,{\bf x}_2)
= \frac{\langle\Psi^\dag({\bf x}_1)\Psi^\dag({\bf x}_2)\Psi({\bf x}_1)\Psi({\bf x}_2)\rangle}
{\langle\Psi^\dag({\bf x}_1)\Psi({\bf x}_1)\rangle \ \langle\Psi^\dag({\bf x}_2)\Psi({\bf x}_2)\rangle}
,
\end{eqnarray}
where $\Psi({\bf x})$ is the bosonic field operator that annihilates a molecule at position ${\bf x}$. The probability to find two particles at the same position ${\bf x}$ is proportional to $g^{(2)}({\bf x},{\bf x})$. For a homogeneous system, this quantity is independent of $\bf x$ and we denote it simply as $g^{(2)}$. For an uncorrelated system $g^{(2)}=1$.

\begin{figure}[tb!]
\centering
\includegraphics[width=7cm]{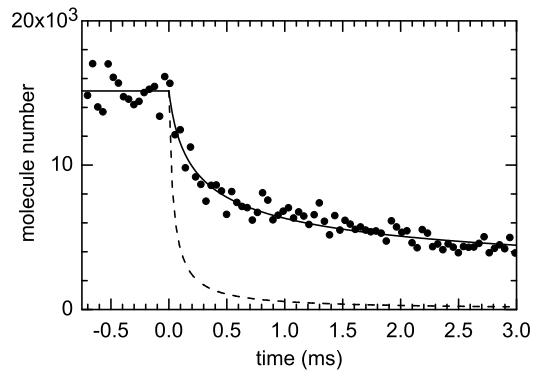}
\caption{\label{fig-decay-tubes}
Time-resolved loss of the number of molecules in 1D tubes. If the system were uncorrelated the loss would be expected to follow the dashed line, which is way off the experimental data ($\bullet$). A fit to the data (solid line) reveals that the probability to find two particles at the same position is reduced by a factor of $\sim 10$ compared to an uncorrelated system, thus showing that the system is strongly correlated. Reproduced from Ref.~\cite{syassen:08}.}
\end{figure}

Loss of particles due to inelastic two-body collisions occurs only if the particles come close together. The rate at which the loss occurs thus depends on $g^{(2)}$; more quantitatively \cite{syassen:08}
\begin{eqnarray}
\frac{d n}{dt} = - K n^2 g^{(2)}
,
\end{eqnarray}
where $n$ is the 1D density of particles and $K$ is a rate coefficient, which can be determined from independent measurements. A measurement of the loss rate can thus serve as a probe whether the strongly correlated regime is reached.

The following experimental procedure is used to study this effect: First, a state with exactly one molecule at each lattice site is prepared as in Ref.\ \cite{volz:06}. Second, the lattice depth along one direction is lowered to zero in 0.5 ms. Third, the system is allowed to evolve for a variable hold time, during which the relevant loss occurs, and finally, the molecule number is measured. 

Figure \ref{fig-decay-tubes} shows experimental data ($\bullet$) for the decay of the molecule number as a function of the hold time. No noticeable loss is observed during the lattice ramp down, which ends at $t=0$. The following loss is way off the expectation for an uncorrelated system (dashed line), which is calculated from the independently determined parameters of the system, including a measurement of the 3D loss rate coefficient in Ref.\ \cite{syassen:06}. The solid line shows a fit to the data that reveals a value of $g^{(2)}=0.11\pm0.01$, see Ref.\ \cite{syassen:08} for details. The fact that $g^{(2)}$ differs from 1 by a large factor shows that the system is strongly correlated, thus realizing a dissipative Tonks-Girardeau gas.

\begin{figure}[tb!]
\centering
\includegraphics[width=4.5cm]{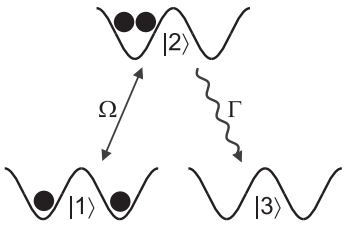}
\caption{\label{fig-Zeno}
Understanding the loss in terms of the quantum Zeno effect. The initial state $|1\rangle$ contains exactly one particle at each site of a double-well potential. Tunneling with amplitude $\Omega$ coherently couples this state to state $|2\rangle$, where both particles occupy the same site. In this configuration, the particles can collide inelastically, resulting in loss of both particles, thus transferring the system into state $|3\rangle$. The rate coefficient for this incoherent loss is $\Gamma$. In the limit $\Omega\ll\Gamma$, loss from the initial state occurs at an effective rate $\Gamma_{\rm eff}=\Omega^2/\Gamma$ \cite{cohen-tannoudji:92:p49}. If $\Gamma$ is large, then $\Gamma_{\rm eff}$ becomes small. Fast dissipation thus freezes the system in its initial state, which can be interpreted as a manifestation of the continuous quantum Zeno effect \cite{misra:77}. Reproduced from Ref.~\cite{syassen:08}.
 }
\end{figure}

An interesting variation of this experiment is obtained when considering the situation where the lattice depth $V_\parallel$ along the 1D tubes is lowered to a nonzero value. Of course, this is closely related to the above experiment, but there are three aspects that make this system interesting: first, the case $V_\parallel\neq0$ offers new physical insight because the reduction of the loss can be interpreted in terms of the quantum Zeno effect as illustrated in Fig.\ \ref{fig-Zeno}, second, time-resolved calculations of the dynamics of the loss become numerically feasible, and third, a much larger suppression of $g^{(2)}$ is obtained.

\begin{figure}[tb!]
\centering
\includegraphics[width=7cm]{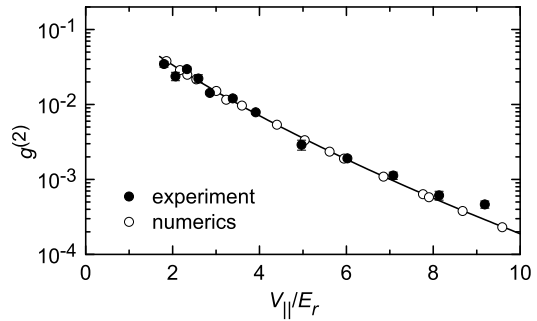}
\caption{\label{fig-g2-lattice}
Pair-correlation function as a function of the lattice depth applied along the one dimension. Experimental data ($\bullet$), numerical results ($\circ$), and analytical model (sold line) agree well with each other. Reproduced from Ref.~\cite{syassen:08}.
}
\end{figure}

In the following, we concentrate on the last two aspects. The pair-correlation function can again be determined from time-resolved measurements of the loss of molecule number. Results ($\bullet$) are shown in Fig.\ \ref{fig-g2-lattice} as a function of $V_\parallel/E_r$, where $E_r$ is the molecular recoil energy. The solid line shows an analytical model discussed in Ref.\ \cite{syassen:08} that represents essentially the Zeno effect illustrated in Fig.\ \ref{fig-Zeno}. In addition, we performed time-resolved numerical calculations that make much fewer approximations than the analytical model. The numerical results are also shown in Fig.\ \ref{fig-g2-lattice} and agree well with the analytical model and the experimental data. The lowest value of $g^{(2)}$ measured here is $\sim 1/2000$.

To summarize, inelastic collisions can be used to drive a many-body system into the strongly correlated regime, much like elastic collisions. This general concept is illustrated in an experimental realization of a dissipative Tonks-Girardeau gas. The suppression of the loss rate due to the correlations is used to measure the pair-correlation function, which quantifies the degree of correlation in the experiment. The physical origin of the suppression can be interpreted in terms of index mismatch in Fresnel's formula or in terms of the quantum Zeno effect.

We acknowledge financial support of the German Excellence Initiative via the program Nanosystems Initiative Munich and of the Deutsche Forschungsgemeinschaft via SFB~631. JJGR acknowledges financial support from the Ramon y Cajal Program and the Spanish projects FIS2006-04885 and CAM-UCM/910758.

\end{document}